\documentclass[twocolumn,prb,aps,psfig,showpacs,preprintnumbers,superscriptaddress]{revtex4}
\usepackage{graphicx}
\usepackage{epstopdf}
\usepackage{float}
\usepackage{color}
\usepackage{amsmath}

\begin{document}
\title{Review of  NMR studies of nanoscale molecular magnets composed of geometrically frustrated antiferromagnetic triangles }

\author{Yuji Furukawa}
\affiliation{Ames Laboratory, U.S. DOE,  Iowa State University, Ames, Iowa 50011, USA}
\affiliation{Department of Physics and Astronomy, Iowa State University, Ames, Iowa 50011, USA}

\date{\today}

\begin{abstract} 
   We present a comprehensive review of nuclear magnetic resonance (NMR) studies performed on three nanoscale molecular magnets with different configurations of geometrically frustrated antiferromagnetic (AFM) triangles, new spin frustration systems with different novel structures: 
(1) the isolated single AFM triangle  K$_6$[V$_{15}$As$_6$O$_{42}$(H$_2$O)]$\cdot$8H$_2$O (in short V15), 
(2) the spin ball  [Mo$_{72}$Fe$_{30}$O$_{252}$(Mo$_2$O$_7$(H$_2$O))$_2$ (Mo$_2$O$_8$H$_2$(H$_2$O))(CH$_3$COO)$_{12}$(H$_2$O)$_{91}$]$\cdot$150H$_2$O (in short Fe30 spin ball), and (3) the twisted triangular spin tube [(CuCl$_2$tachH)$_3$Cl]Cl$_2$ (in short Cu3 spin tube).
   In the V15 nanomagnet, from $^{51}$V NMR spectra observed at only below $\sim$ 0.1 K, we directly determined the local spin configuration in both the nonfrustrated total spin $S_{\rm T}$ = 3/2 state at higher magnetic fields ($H \geq 2.7 $ T)  and the two nearly degenerate $S_{\rm T}$ = 1/2 ground states at lower magnetic fields ($H \leq 2.7 $ T).  
   The dynamical magnetic properties of V15 were investigated by  proton spin-lattice relaxation rate (1/$T_1$) measurements. 
    In the $S_{\rm T}$ = 3/2 state,   1/$T_1$ shows thermally activated behavior as a function of temperature. 
    On the other hand, a temperature independent behavior of 1/$T_1$ at very low temperatures is observed in the frustrated $S_{\rm T}$ = 1/2  ground state below 2.7 Tesla. 
    Possible origins for the peculiar behavior of 1/$T_1$ will be discussed in terms of magnetic fluctuations due to spin frustrations. 
  In Fe30 spin ball, static and dynamical properties of Fe$^{3+}$ ($s = 5/2$) have been investigated by proton NMR spectra and 1/$T_1$ measurements. 
   From the temperature dependence of 1/$T_1$, the fluctuation frequency of the Fe$^{3+}$ spins is found to decrease with decreasing temperature, indicating spin freezing at low temperatures. 
  The spin freezing is also evidenced by the observation of a sudden broadening of $^1$H NMR spectra below 0.6 K. 
  Finally $^1$H NMR data in the Cu spin tube will  be described. 
  An observation of magnetic broadening of $^1$H NMR spectra at low temperatures below 1 K directly revealed a gapless ground state of the system. 
   The $1/T_1$ measurements revealed an usual slow spin dynamics in the quasi one dimensional Cu3 spin tube.

\end{abstract}


\maketitle

\section{INTRODUCTION}

    $``$Spin frustration$"$ is one of the most exciting frontiers of contemporary research in condensed matter physics.\cite{Diep, Lacroix} 
     Historically, the interest in spin frustration was raised long ago by Anderson’s prediction for a resonating valence bond (RVB) state (sometimes called $``$spin liquid state$"$), as an alternative to the classical N\'eel state \cite{Anderson1973} for a magnetic ground state in a two-dimensional triangular-lattice $s$ = 1/2 antiferromagnet. 
   From a classical point of view,\cite{Schender1994,Reimers1993}  this generates a very high degeneracy of the ground state, translating into a huge density of low energy excitations, no long-range order at $T$ = 0 K, and a very short magnetic correlation length characteristic of a $``$spin liquid state$"$. 
     Experimentally, many of the spin frustrated systems, such as triangular-lattice, Kagom\'e-lattice and pyrochlores, have been investigated to elucidate the magnetic ground state. 
     However, most of the systems show short-range or long-range magnetic order, due to small perturbations such as defects, anisotropy, and lattice distortion.\cite{Inami2001}
    This prevents the effects purely associated with spin frustrations from direct observation. 
    Therefore, it is difficult to clarify what kind of state is actually realized as the ground state of such a strongly frustrated system and more ideal compounds to investigate the effects of spin frustration in the ground state are strongly desired.

\begin{figure}[b]
\includegraphics[width=9.0cm]{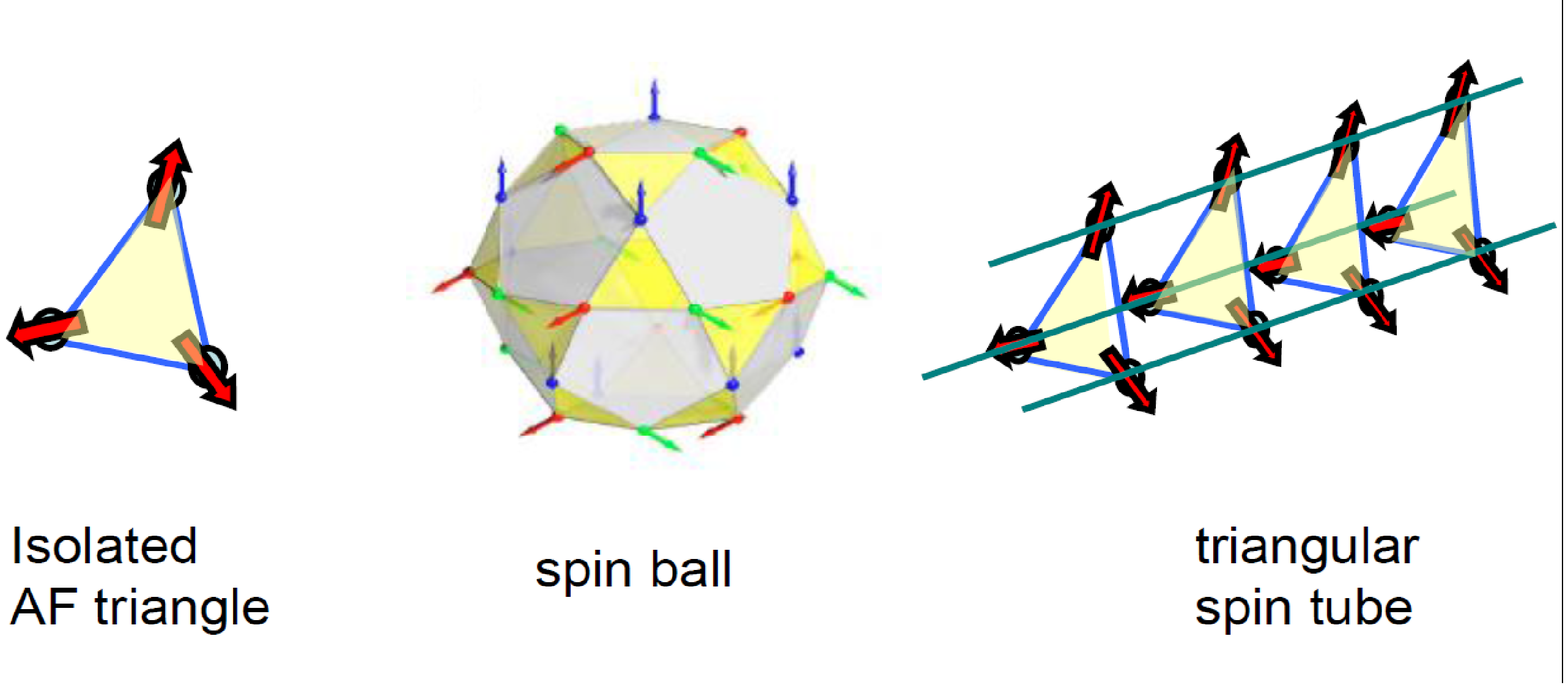}
	\caption{(Color online) New spin systems with novel structures. Each arrow represents spin on the magnetic ion. }
	\label{fig:Fig.1}
\end{figure}

       Recent progress in synthesizing nanoscale molecular magnets offers the opportunity to investigate the effects of spin frustration at nanoscale levels.\cite{Gatteschi1991,McInnes2005} 
      Molecular magnets are composed of a controllable number of magnetic ions with spins located on a triangle, tetrahedron, ring, ball, and other geometries.\cite{Gatteschi2006} 
    Since the distance between the magnetic molecules themselves is usually more than $\sim$ 1.5 nm, the magnetic interaction between the molecules is very small and usually three-dimensional magnetic ordering does not occur even at very low temperatures down to mK. 
    Thus, one can investigate the magnetic properties of the single molecular system at very low temperatures without any disturbance from magnetic orderings. 
    The simplest spin frustrated system is three  $s$ = 1/2 antiferromagnetically coupled spins with a triangular configuration. 
     In a molecular nanomagnet, this simplest spin frustration system can really exist and one can investigate this simplest system experimentally.\cite{Gatteschi1991}

    An innovative idea to utilize the nanoscale molecular magnet as a magnetic unit produced a great breakthrough for synthesizing a new class of spin frustration systems called spin ball \cite {Muller1999,Schroder2005,Lago2007}  and triangular spin tube.\cite{Seeber2004,Schnack2004}
     The spin ball system has a spherical shape constructed from twenty AFM triangles with spin frustration. 
      In the spin tube, AF triangles form a one-dimensional chain as shown in Fig. 1. 
      Although these peculiar structures have attracted much interest \cite{Kawano1997,Luscher2004,Fouet2005,Okunishi2005, Sato2007_2,Sato2007,Nishimoto2008,Sakai2008,Sakai2010,Lajko2012}
 the systems were considered to be only toy models for theoretical study, due to the lack of real model materials before the synthesis breakthrough.

     In this review, we discuss the magnetic properties of these peculiarly-structured spin frustration systems studied by using nuclear magnetic resonance (NMR) technique.
      NMR has proved to be a powerful tool for investigating both static and dynamic properties of nanoscale molecular magnets.\cite{Borsa2006, Borsa2008} 
     The NMR spectrum gives us information on the hyperfine interaction of the nuclei with the local magnetic moments, in particular  the $``$on-site$"$  NMR spectrum of nuclei on magnetic ions provides direct information on the  local magnetic moments of the magnetic ions.
    On the other hand, the nuclear spin lattice relaxation rate (1/$T_1$) directly probes the low-frequency spectral weight of spin fluctuations of the local magnetic moments.

\section{EXPERIMENTAL DETAILS}

      The polycrystalline powder samples were synthesized as described in Ref. \onlinecite{Choi2003} for V15, Ref. \onlinecite{Muller1999} for Fe30 and Ref. \onlinecite{Seeber2004} for the Cu spin tube. 
      The NMR measurements were performed utilizing  a  homemade phase-coherent spin-echo pulsed NMR spectrometer on $^{1}$H  nuclei (nuclear spin $I$ = 1/2 and gyromagnetic ratio $\gamma$$_{\rm N}$/2$\pi$ = 42.5774 MHz/T)  and $^{51}$V nuclei ($I$ = 7/2 and $\gamma$$_{\rm N}$/2$\pi$ = 11.193 MHz/T)  in the $T$ range 0.1 K $\leq$ $T$ $\leq$ 300 K using a $^{3}$He-$^{4}$He dilution refrigerator (Kelvinox MX100, Oxford instruments) installed at the Ames Laboratory by the author. 
      A part of NMR data was taken at Hokkaido University where   a $^{3}$He-$^{4}$He dilution refrigerator (Kelvinox100, Oxford instruments) was also installed by the author and collaborators.      
     The NMR spectrum was obtained either by Fourier transform of the NMR echo signal at a constant magnetic field or by sweeping the magnetic field. 
    The NMR echo signal was obtained by means of a Hahn echo sequence with a typical $\pi$/2 pulse length of 1.5 $\mu$s.
    The nuclear spin-lattice relaxation rate 1/$T_1$ was measured by the saturation method at the peak position of the NMR spectrum. 
    The nuclear magnetization recoveries for $^1$H were found to be slightly non exponential due to the existence of protons with inequivalent spatial locations. 
    Such non-exponential behavior has been reported in many magnetic molecules.\cite{Borsa2006} 
    $1/T_1$ values were determined from the initial slope of the recovery behavior, which corresponds to a weighted average relaxation rate of the nonequivalent nucleus.\cite{Borsa2006}

\section{V15: an isolated triangular antiferromagnet }
      K$_6$[V$_{15}$As$_6$O$_{42}$(H$_2$O)]$\cdot$8H$_2$O (in short, V15) is one of the interesting molecular magnet for investigating spin frustration since V15 can be  considered a model system for an $s=1/2$ Heisenberg single triangle antiferromagnet, the simplest spin frustration system.\cite{Gatteschi1991} 
   V15 is comprised of fifteen V$^{4+}$ ions with $s=1/2$, which are arranged in a 
quasi-spherical layered structure with a triangle sandwiched between two hexagons as shown 
in Fig. 2(a). 
   All exchange interactions between V$^{4+}$ spins are antiferromagnetic (AF).\cite{Gatteschi1993,Barra1992} 
   Each hexagon of V15 consists of three pairs of strongly coupled spins with $J_1$ $\sim$ 800 K. 
   Each spin of V$^{4+}$ ions in the central triangle is coupled with the spins in both hexagons 
with $J_2$ = 150 K and $J_3$ = 300 K,\cite{Chaboussant2002} resulting in a very weak exchange interaction between the 
spins within the central triangle with $J_0$ = 2.44 K.\cite{Chaboussant2002} 
   At low temperatures, the magnetic properties of V15 are  determined entirely 
by the three V$^{4+}$ spins on the triangle (a frustrated $s=1/2$ triangular system) 
because the V$^{4+}$ spins on the hexagon are assumed to form a total $S_{\rm T}$ = 0 spin singlet state due 
to the strong AF interaction of $J_1$ $\sim$ 800 K, 
as confirmed directly by $^{51}$V-NMR measurements at very low temperatures.\cite{Nishisaka2006}

\begin{figure}[b]
\includegraphics[width=8.5cm]{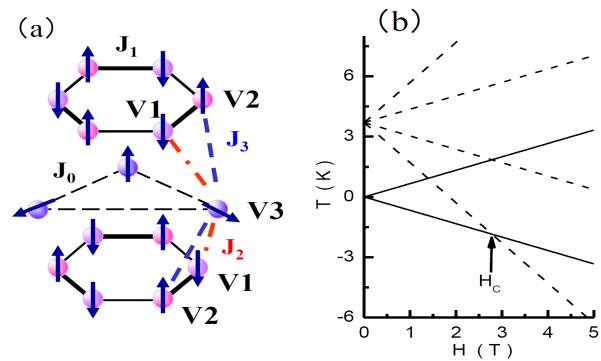}
	\caption{(Color online) (a) Schematic view of the relative positions of V$^{4+}$ ($s=1/2$) ions 
(solid circles) and the exchange coupling scheme in the V15. 
(b) Magnetic energy level scheme of the V15 as a function of the external magnetic field. 
Solid and broken lines show the nearly degenerated two $S_{\rm T}$ = 1/2 branches and the 
$S_{\rm T}$ = 3/2 branches, respectively. }
	\label{fig:Fig.2}
\end{figure}

   The magnetic levels of V15 at low temperature can thus be described by a 
simple model spin Hamiltonian
\begin{eqnarray}
H 
& = & (J_{12} {\bf S}_1 \cdot {\bf S}_2 + J_{23}{\bf S}_2 \cdot {\bf S}_3 
+J_{31}{\bf S}_3 \cdot {\bf S}_1) \nonumber \\
& + & g \mu_{B} {\bf H} \cdot ({\bf S}_1 + {\bf S}_2 + {\bf S}_3)
\eqnum{1}
\label{eqn:Spin_Ham}
\end{eqnarray}
where the exchange parameters between spins can be assumed equal in first approximation, i.e.,
$J_{12}$ = $J_{23}$ = $J_{31}$ = $J_0$ and $S_i$ are the individual $s=1/2$ V sites. 
     The energy scheme at low temperature is given by a ground state of two degenerate total 
spin $S_{\rm T}$ = 1/2 and an excited state of $S_{\rm T}$ = 3/2 which lies $\sim$ 3.8 K above the ground state.\cite{Dobrovitski2000} 
   In reality, the two $S_{\rm T}$ = 1/2 states are split with a small energy gap which was estimated to 
be of the order of 0.08 K from the magnetization measurement.\cite{Chiorescu2000} 
   Among the possible origins for the gap have been discussed in terms of  a small Dzyaloshiskii-Moriya (DM) interaction,\cite{Chiorescu2000_2,Miyashita2001,Raedt2004} 
hyperfine interactions\cite{Miyashita2003} and lattice distortion.\cite{Chaboussant2004}

\subsection{$^{51}$V NMR spectra in V15}

\begin{figure}[b]
\includegraphics[width=8.0cm]{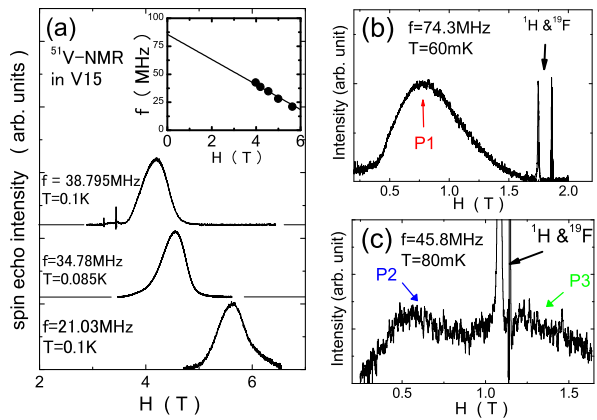}
	\caption{(Color online) (a) $^{51}$V-NMR spectra for the V$^{4+}$ ions in the triangle observed in 
its $S_{\rm T}$=3/2 ground states at very low temperature for three different resonance frequencies. 
The inset shows the external field dependence of resonance frequency for the peak measured  
below 100 mK. 
   Typical $^{51}$V-NMR spectra observed in  the $S_{\rm T}$=1/2 ground states are shown in (b) and (c).}
	\label{fig:Fig.3}
\end{figure}

    $^{51}$V NMR signal was observed successfully only below 0.1 K.\cite{Nishisaka2006,Furukawa2007} 
    Above $H$ = 2.7 T,  where the ground state of the V15 cluster is $S_{\rm T}$ = 3/2, a single peak of $^{51}$V NMR spectrum with full width of half maximum (FWHM) of $\sim$ 5 kOe is observed. 
    The peak position of the spectrum shifts to lower magnetic field with increasing resonance frequency as shown in Fig. 3(a). 
    The resonance frequency $f$ is proportional to the vector sum of the internal field $H_{\rm int}$ and external field $H_0$: 
\begin{eqnarray}
f = \frac{\gamma_{\rm N}}{2\pi} ( H_{\rm 0} + H_{\rm int} )
\label{eqn:res_freq}
\end{eqnarray}
where $\gamma_{\rm N}$/2$\pi$ = 11.285 MHz/T is the gyromagnetic ratio of the $^{51}$V 
nucleus. 
     By fitting the data points in the field above 2.7 T, as shown in the inset of Fig. 3(a), $H_{\rm int}$ is estimated to be -- 7.6 T. 
   The internal field at the nuclear site in a V$^{4+}$ ion with $s$ = 1/2 is dominated by the core polarization mechanism which induces a large negative internal field of the order of 100 kOe/$\mu_{\rm B}$ at the nuclear site.\cite{Watoson1967} 
   The value of -- 76 kOe is close to the value of -- 85 kOe reported in VO$_2$.\cite{Takahashi1983} 
    Thus this $^{51}$V signal can be assigned to V$^{4+}$  ions with an almost full spin moment of 1 $\mu_{\rm B}$. 
      In addition, the good fitting result shown by solid line in the inset of Fig. 3(a) indicates the direction of the spin moments is parallel to the external field. 
     Thus we conclude that each V$^{4+}$ ion on the triangle has almost the full spin moment of 1 $\mu_{\rm B}$ whose direction is parallel to the external field, which gives the total spin $S_{\rm T}$ = 3/2  state for the ground state of the V15 cluster above 2.7 T. 
     This is microscopic evidence of the spin structure of the V15 cluster for the $S_{\rm T}$ = 3/2  ground state.

\begin{figure}[b]
\includegraphics[width=7.5cm]{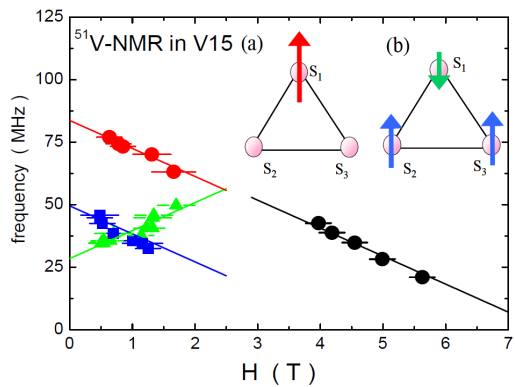}
	\caption{(Color online) External magnetic field dependence of the resonance frequency for the three $^{51}$V-NMR 
signals observed in $S_{\rm T}$ = 1/2 ground states at $H_{\rm 0}$ $<$ 2T.  
    The solid circles represent the stronger signal (P1) shown in Fig. 3(b). 
     The solid squares and triangles refer to the other two weaker signals P2 and P3, respectively, 
shown in Fig. 3(c). 
    The high field data in Fig. 3(a) have been replotted here for comparison. 
    The solid lines show the fitting results using Eq. (2). 
    The inset shows schematic view of the expectation values for the spin moment for $S_1$, $S_2$ and $S_3$ 
sites on the triangle according to the eigenfunctions : (a) $\psi_{\rm \alpha}$,  
(b) $\psi_{\rm \beta}$.}
	\label{fig:Fig4}
\end{figure}

      On the other hand, three different V sites (P1, P2 and P3) are observed in the case of $S_{\rm T} =1/2$ ground state below 2.7 T as shown in Fig. 3(b) and (c). 
     The external field dependence of those peak positions are plotted as circles, squares and triangles in Fig. 4. 
    By fitting these data using Eq. 2, internal fields for three V sites are estimated to be -- 7.5 T, -- 4.5 T and 2.5 T for P1, P2 and P3, respectively. 
     Assuming the hyperfine field for V$^{4+}$ ions in V15 due to the core polarization is -- 7.6 T for $s$ = 1/2, spin moments for each each V$^{4+}$ site  can be estimated to be $\sim$ 1 $\mu_{\rm B}$, 0.6 $\mu_{\rm B}$ and -- 0.33 $\mu_{\rm B}$ for P1, P2 and P3, respectively, where the positive and negative signs of the spin moments correspond to the direction of parallel and antiparallel with respect to the external field, respectively. 

    Now we compare the experimental findings with the theoretical predictions. 
     The two eigenfunctions  for the $s=1/2$ Heisenberg triangular system with spin Hamiltonian Eq. 1 
corresponding to the eigenstate given by the two $S_{\rm T}$=1/2 ground states can be expressed using basis functions of 
$\mid$$S_1 S_2 S_3$$\rangle$ in which up and down arrows represent up and down spin, respectively, for $S_n$ ;
\begin{eqnarray}
\phi_{\rm a} 
& = & \frac{1}{\sqrt{3}} (\mid \uparrow \downarrow \downarrow \rangle 
               +  \omega \mid \downarrow \uparrow \downarrow \rangle
               +  \omega^{2} \mid \downarrow \downarrow \uparrow \rangle) \nonumber \\
\phi_{\rm b} 
& = & \frac{1}{\sqrt{3}} (\mid \uparrow \downarrow \downarrow \rangle 
               +  \omega^{2} \mid \downarrow \uparrow \downarrow \rangle
               +  \omega \mid \downarrow \downarrow \uparrow \rangle)   \eqnum{3}
\label{eqn:eigen_chiral}
\end{eqnarray}                                                 
where $\omega$=e$^{2 \pi i/3}$. 
    These two eigenfunctions correspond to two different spin chiral states in which V$^{4+}$ spin rotates 120 degrees 
with respect to neighboring spins respectively in clockwise and counterclockwise directions on the triangle. 
    Other eigenfunctions for the same Hamiltonian can be obtained by making linear combinations of the basis functions. 
    The two functions below are also eigenfunctions for the $S_{\rm T}$=1/2 frustrated spin state:
\begin{eqnarray}
\psi_{\rm \alpha}
& = & \frac{1}{\sqrt{2}} (\mid \downarrow \downarrow \uparrow \rangle
                                       - \mid \downarrow \uparrow \downarrow \rangle) \nonumber \\
\psi_{\rm \beta}
& = & \frac{1}{\sqrt{6}} (2\mid \uparrow \downarrow \downarrow \rangle 
               - \mid \downarrow \uparrow \downarrow \rangle
               - \mid \downarrow \downarrow \uparrow \rangle) .   \eqnum{4}
\label{eqn:eigen_V15}
\end{eqnarray}
    Although the total spin moment for both sets of eigenfunctions (Eq. 3 and Eq. 4) is  $S_{\rm T}$ = 1/2 
(1 $\mu_{\rm B}$) as required, the expectation values for local spin moments for $S_1$, $S_2$ and $S_3$ site 
are estimated to be 1$\mu_{\rm B}$, 0$\mu_{\rm B}$ and 0 $\mu_{\rm B}$ for the $\psi_{\rm \alpha}$ 
wavefunction  and  -1/3 $\mu_{\rm B}$, 2/3$\mu_{\rm B}$ and 2/3 $\mu_{\rm B}$ for 
the  $\psi_{\rm \beta}$,  respectively, as schematically shown in the inset of Fig. 4. 
    On the other hand, the expectation values for the local spin moments for the  $\phi_{\rm a}$ and 
$\phi_{\rm b}$  eigenfunctions in Eq. 3 are calculated to be 1/3 $\mu_{\rm B}$ for each spin. 
   Since experimentally three V$^{4+}$ ions in the $S_{\rm T}$ = 1/2 ground state with different spin moments 
(1 $\mu_{\rm B}$, -- 1/3 $\mu_{\rm B}$ and 2/3 $\mu_{\rm B}$) are observed, we conclude that the  
two eigenfunctions to be associated with the  two $S_{\rm T}$ = 1/2 
ground states can be expressed by $\psi_{\rm \alpha}$ and  $\psi_{\rm \beta}$.

   Each of the two eigenfunctions should be assigned to one of the two quasidegenerate 
$S_{\rm T}$=1/2 ground states. 
   Since the measurements were performed at a temperature smaller than the splitting of the 
two quasi-degenerate states the lower energy state is more populated and thus should yield a 
larger NMR signal.
   Since  we observe a larger signal intensity for the $^{51}$V NMR signal arising from a V$^{4+}$ moment of 
1 $\mu_{\rm B}$  (see the inset in Fig. 4) we conclude that the lower energy of the two $S_{\rm T}$ = 1/2 states 
correspond to the eigenfunction  $\psi_{\rm \alpha}$ (see the inset (a) in Fig. 4). 
   The other two weak  $^{51}$V-NMR signals (see Fig. 3(c)) arise from V$^{4+}$ ions with moments 
(-- 1/3 $\mu_{\rm B}$ and 2/3 $\mu_{\rm B}$)  and correspond thus to the eigenfunction  $\psi_{\rm \beta}$ 
(see the inset (b) in Fig. 4) pertaining to the higher energy $S_{\rm T}$=1/2 state. 
   The above conclusions are in agreement with spin densities calculations which indicate  
$\psi_{\rm \alpha}$ as one of the eigenfunctions for the $S_{\rm T}$=1/2 ground state.\cite{Raghu2001} 

\subsection{Spin dynamics in V15}

   Figure 5 shows the temperature dependence of the proton-1$/T_1$ under various magnetic fields.\cite{Furukawa2007_2}
    In magnetic fields above 2.7 T, where the ground state of the cluster is $S_{\rm T} =3/2$, 1$/T_1$ decreases with decreasing temperature at a rate which increases with increasing magnetic field. 
     Below 2.7 T where the ground state of the cluster is the nearly degenerate $S_{\rm T}=1/2$ states, 1$/T_1$ decreases on lowering the  temperature. 
     But, at very low temperatures, 1$/T_1$ shows temperature independent behavior with a constant value that  does not depend on the magnetic field, as shown in Fig. 5.

\begin{figure}[b]
\includegraphics[width=8.0cm]{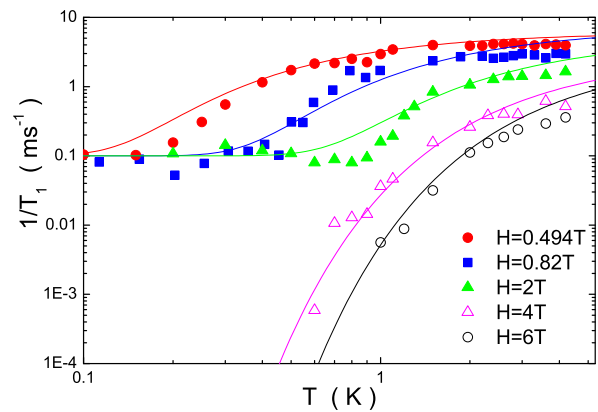}
	\caption{(Color online) Temperature dependence of 1$/T_1$ under various magnetic fields for V15. Solid lines show fitted fitting results discussed in the text. }
	\label{fig:Fig5}
\end{figure}

     Except for the temperature independent behavior of 1$/T_1$ at very low temperatures in the $S_{\rm T}=1/2$ ground states, the temperature dependence of 1$/T_1$ can be explained as thermally activated behavior 
1/$T_1\sim {\rm exp}(-\Delta/k_{\rm B}T$), with a gap magnitude of $\Delta$. 
   The lines in Fig. 5  show fitting results using a relation 1/$T_1 ={\rm exp}(-\Delta/k_{\rm B}T$) + $a$ where $a$  is a constant value of $\sim$ 0.1 (msec$^{-1}$) for the $S_{\rm T}=1/2$ ground states and is zero for the $S_{\rm T}=3/2$ ground state. 
    The $H$ dependence of $\Delta$ estimated from the temperature dependence of 1$/T_1$ follows a relation of $\Delta$~=~1.33$H_{\rm ext}$ (K) where $H_{\rm ext}$  is the external field in Tesla. 
   Since $\Delta$~=~1.33$H_{\rm ext}$ corresponds to the Zeeman energy between the ground state and the first excited magnetic sublevels in the same total spin state, the $T$ dependent behavior of 1$/T_1$ can be explained by thermal fluctuations of the magnetization between the ground and first excited sublevels for each total spin state. 

    On the other hand, the temperature independent behavior of 1$/T_1$ at very low temperatures in the nearly degenerate $S_{\rm T}=1/2$ ground state cannot be explained by thermal fluctuations of the magnetization. 
    Since most of the clusters occupy the nearly degenerate two $S_{\rm T}=1/2$ ground state sublevels at low enough temperature, magnetic fluctuations can be considered only between the two nearly degenerate $S_{\rm T}=1/2$ ground states. 
   The temperature and magnetic field independent behavior of 1$/T_1$ indicates that the magnetic fluctuations between $S_{\rm T}=1/2$ ground states do not depend on temperature and magnetic field.

    As shown by Eq. (3), in the case of an $s=1/2$ antiferromagnetic Heisenberg triangular magnet, the  two $S_{\rm T}=1/2$ states can be regarded as two states with different spin chirality (different spin structures).\cite{Vallain1997,Kawamura1998}
    Since the fluctuations between two chiral states with different spin structures would give rise to a local field fluctuation at the proton sites, such chiral fluctuations could be a relaxation mechanism which gives rise to temperature independent behavior of 1/$T_1$. 
    However, as shown in the Sec. III-A,   the two $S_{\rm T}=1/2$ states in the V15 cluster are described not by the two purely chiral states but by different states given by mixtures of the two chiral states.
    Although it is not clear whether the temperature independent behavior of 1$/T_1$ originates from the chiral fluctuations between the mixed chiral states of the $S_{\rm T}=1/2$ states in the V15 system or not, the temperature and magnetic field independent behavior of 1$/T_1$ apparently relates to the existence of the two $S_{\rm T}=1/2$ states, which originates from the spin frustration.

\section{Fe30 spin ball}
    The compound [Mo$_{72}$Fe$_{30}$O$_{252}$(Mo$_2$O$_7$(H$_2$O))$_2$ (Mo$_2$O$_8$H$_2$(H$_2$O))(CH$_3$COO)$_{12}$(H$_2$O)$_{91}$]$\cdot$150H$_2$O (in short Fe30) is one of the largest molecular paramagnets prepared to date and has 30 Fe$^{3+}$ ($s$  = 5/2) ions occupying the 30 vertices of an icosidodecahedron (see Fig. 1), resulting in  a closed spherical structure of 2.5 nm diameter.\cite{Muller2001} 
    This polyhedron contains 20 corner-sharing Fe3 triangles with antiferromagnetic exchange coupling ($J/k_{\rm B} \sim 1.57$ K) between Fe spins.\cite{Muller2001, Schroder2005} 

\begin{figure}[tb]
\includegraphics[width=8.0cm]{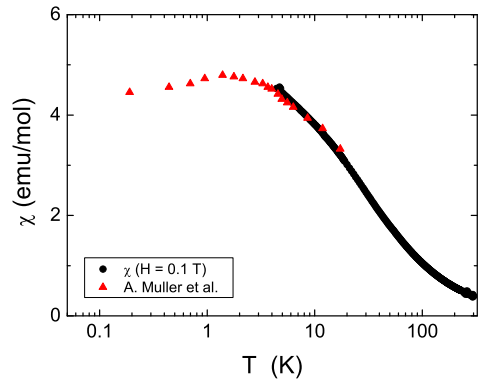}
	\caption{(Color online) (a)  Temperature dependence of  the magnetic susceptibility  measured at $H = $ 0.1 T in a temperature range of 4.0 - 300 K using a superconducting quantum interface device (SQUID) magnetometer (Quantum design MPMS-7T). 
  Data reported by M\"uller {\it et~al.} are shown by red triangles.\cite{Muller2001}
  }
	\label{fig:Fe30chi}
\end{figure}

      The compound shows a simple paramagnetic behavior down to $\sim$ 20 K where the magnetic susceptibility departs from a simple Curie-Weiss law. The magnetic susceptibility shows a broad peak around 2 K and is almost constant below 1 K down to 0.12 K.\cite{Muller2001} 
     No three dimensional magnetic ordering down to 60 mK was reported from magnetization measurements.\cite{Schroder2008}
   Figure 6 shows the temperature dependence of $\chi$ together with data reported by M\"uller {\it et~al}.\cite{Muller2001} 
   The broad peak around 2 K associated with the short-range ordering due to the antiferromagnetic interaction $J/k_{\rm B} \sim$ 1.57 K between Fe$^{3+}$ spins in the molecules.

\subsection{$^{1}$H NMR spectra in Fe30}

   Figure  7(a) shows $^1$H-NMR spectra at $f$ = 7.03 MHz for several temperatures. \cite{Furukawa2012}
   Figure 7(b) shows the temperature dependence of the full width at half maximum (FWHM) of the $^1$H NMR spectrum at $f = 7.03$ MHz ($H$ = 0.16 T), together with data at  $f = 26$ MHz ($H = 0.61$ T).\cite{Schroder2010} 
     A dramatic broadening of the spectrum is clearly observed at $\sim$ 600 mK for both magnetic fields. 
    The FWHM is almost independent of magnetic field below 600 mK, which indicates freezing state of Fe$^{3+}$ spins. 
     As can be seen in Fig. 7(a), the $^1$H-NMR signal at $ T = 0.1$ and 0.19 K can be observed at $H =$ 0 T, which is also direct evidence of very slow fluctuations of the Fe spin moments. 
     Almost no signal was observed around 1 K due to a shortening of the  nuclear spin-spin relaxation time $T_2$ which follows a corresponding shortening of the nuclear spin-lattice relaxation time $T_1$. 

\begin{figure}[b]
\includegraphics[width=8.0cm]{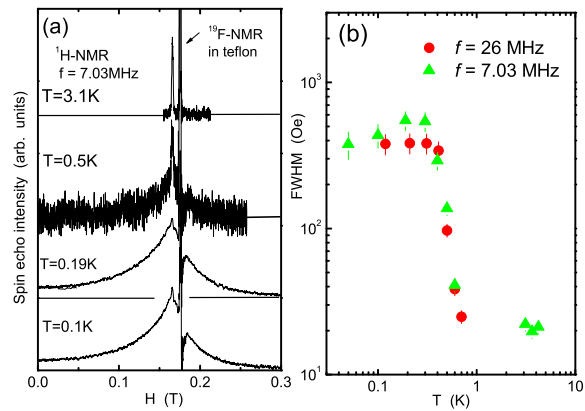}
	\caption{(Color online) (a)  Temperature dependence of $^1$H-NMR spectrum at $f =$ 7.03 MHz.
       (b) Temperature dependence of  FWHM  of $^1$H NMR spectrum in Fe30 at $H =$ 0.61 T (circles) and $H =$ 0.16 T (triangles).}
	\label{fig:Fig6}
\end{figure}

\begin{figure*}[tb]
\includegraphics[width=17.0cm]{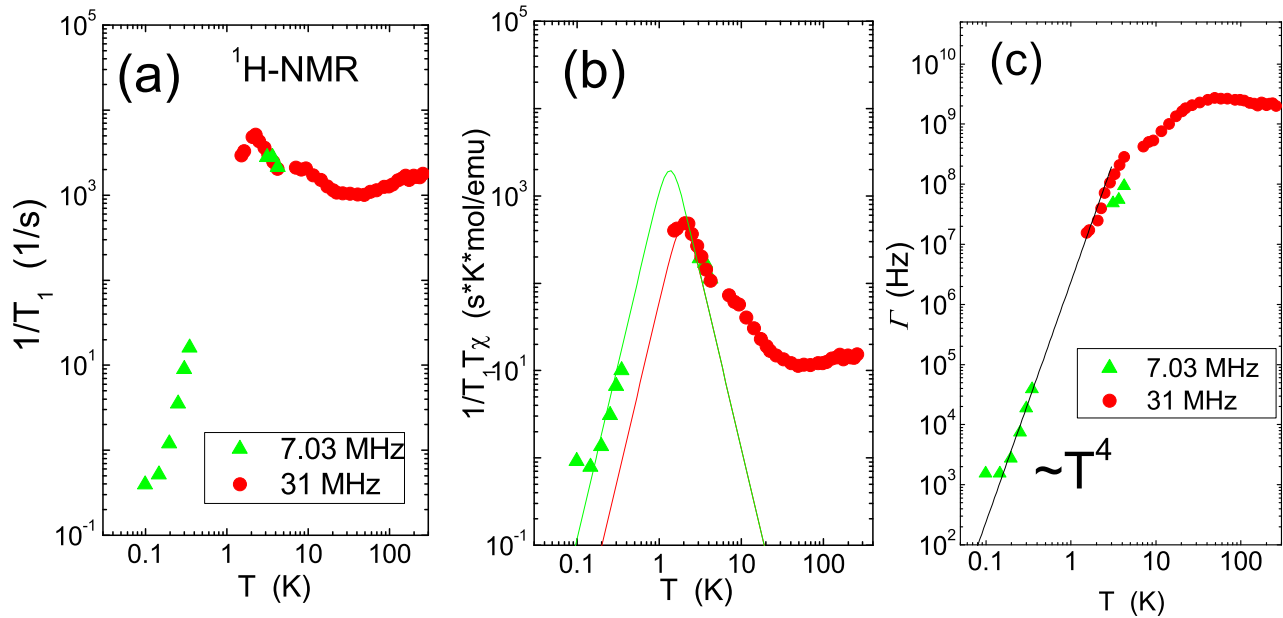}
	\caption{(Color online) (a) Temperature dependence of  proton 1$/T_1$ in Fe30 at $f =$ 7.03 MHz ($H =$ 0.16 T, closed triangles) and 31 MHz ($H = $ 0.75 T, closed circles). 
    (b) Temperature dependence of  1/$T_1T\chi$.  Solid lines are calculated results using Eq. (6). 
    (c) Temperature dependence of $\Gamma$, fluctuation frequency of the Fe spins.  }
	\label{Fig7}
\end{figure*}

\subsection{Spin dynamics in Fe30}

   The spin freezing of Fe$^{3+}$ spins can also be observed in proton 1/$T_1$ measurements.
    Figure 8(a) shows the temperature dependence of 1$/T_1$ under two magnetic fields. 
     With decreasing temperature, 1$/T_1$ for $H =$ 0.75 T decreases gradually and starts to increase around 20 K, then shows a peak around 2 K. 
    Similar temperature dependence of 1$/T_1$ was reported by Lago {\it et~al.},\cite{Lago2007} and the peak of 1$/T_1$ is found to increase in magnitude and move toward low temperature by decreasing magnetic field. 
     The temperature dependence of 1$/T_1$ at $H =$ 0.16 T is also plotted in the figure. 
     1$/T_1$ could not be measured between 0.6 K and 1.5 K because of shortening of $T_2$ (i.e., $T_1$). 
     
      In general,   1/$T_1$ is  expressed by the Fourier transform of the time correlation function of the transverse component $\delta$$h_{\pm}$ of the fluctuating local field at nuclear sites with respect to the nuclear Larmor frequency $\omega_{\rm N}$ as,\cite{Abragam}
 \begin{eqnarray}
\frac{1}{T_1} 
& = & \frac{1}{2} \gamma_{\rm N}^2 \int_{-\infty}^{+\infty} \langle\ h_{\pm}(t) h_{\pm}(0)\ \rangle 
{\rm exp}(i \omega_{\rm N}t) dt  
\label{eqn:$T_1$}
\end{eqnarray}
where $\gamma_{\rm N}$  is the gyromagnetic ratio of the nuclear spin.
   When the time correlation function is assumed to decay as exp(-$\Gamma t$),  one can write\cite{Borsa2006} 
\begin{eqnarray}
\frac{1}{T_1T \chi_0}   =   A  \frac{\Gamma}{\Gamma^2 + \omega_{\rm N}^2} ,
 \label{eqn:$T_1$_6}
\end{eqnarray}
where $A$ and is a parameter related to the hyperfine field at nuclear sites. 
   $\Gamma$ corresponds to the inverse of the correlation time of the fluctuating hyperfine fields at the H sites, due to the Fe$^{3+}$ spins, and thus can be regarded as the fluctuation frequency of the spins.

     To analyze the temperature dependence of 1$/T_1$ by using Eq. (6), it is useful to re-plot the data by changing the vertical axis from 1$/T_1$ to 1/$T_1T\chi$  as shown in Fig. 8(b), where the $\chi$ values for $T = 0.1-4$ K are obtained from $\chi$ data reported by M\"uller in Ref. \onlinecite{Muller2001} and for $T = 4-300$ K we used our data. 
    If $\Gamma$ is independent of temperature, 1/$T_1T\chi$ should be constant which is in fact observed above 30 K in Fig. 7(b). 
     This indicates that the nuclear spin relaxation above $\sim$ 30 K is explained by paramagnetic fluctuations of the Fe$^{3+}$ spins whereby the spins fluctuate almost independently from each other. 
    On the other hand, below 30 K 1/$T_1T\chi$ increases with decreasing temperature. 
   This indicates the simple paramagnetic fluctuations model cannot explain the $T$  dependence of 1/$T_1T\chi$.

      According to Eq. (6), 
1/$T_1T\chi$  is proportional to 1/$\Gamma$  when $\Gamma$ $\gg$ $\omega_{\rm N}$  (fast-motion regime), while 1/$T_1T\chi$ is proportional to $\Gamma$/$\omega_{\rm N}^2$ in the case of $\Gamma$ $\ll$ $\omega_{\rm N}$ (slow-motion regime). 
   When $\Gamma$ = $\omega_{\rm N}$,  1/$T_1$ reaches a maximum value. 
    Thus, the slowing down of the fluctuation frequency $\Gamma$ of Fe$^{3+}$ spins produces a peak of 1/$T_1$. 
 Assuming $A = 1.7 \times 10^{11}$ (rad$^2$$\cdot$K$\cdot$emu/mol/s$^2$) and $\Gamma$ = 1.34 $\times$ 10$^7$ $T^4$ (rad/s), the experimental results at low temperature regions below 4 K are qualitatively reproduced by Eq. (6), as shown in Fig. 8(b) by solid lines for different magnetic fields. 
     These results indicate that the peak observed in the temperature dependence of 1/$T_1T\chi$  originates from a crossover between the fast-motion regime and the slow-motion regime, whereby the fluctuation frequency of Fe$^{3+}$ spins below the peak temperature is less than the NMR frequency range which is of the order of MHz.

     To extract the temperature dependence of the fluctuation frequency of Fe$^{3+}$ spins for a wide temperature region, we estimate the temperature dependence of $\Gamma$ from the temperature dependence of 1/$T_1T\chi$  assuming Eq. (6) is valid for the entire  temperature region. 
     The estimated temperature dependences of $\Gamma$ for the two different magnetic fields are shown in Fig. 8(c). 
     This log-log plot shows very clearly that $\Gamma$ has a power low behavior ($T^{\sim4}$) at low temperatures below $\sim$ 1 K and is nearly constant at $\sim3\times10^9$ Hz at high temperatures. 
      At low enough temperature Fe$^{3+}$ spins can fluctuate with low frequency which is less than the NMR frequency of the order of MHz. 
      Such a slow spin dynamics is consistent with the observation of broadening of NMR spectrum below 600 mK. 

\begin{figure}
\includegraphics [width=8.0cm]{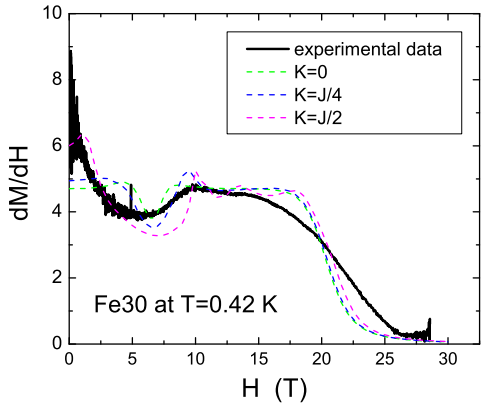}
\caption{(Color online) Simulated d$M$/d$H$ in Fe30 at $T= 0.42$ K. 
    The experimental data from Ref. \onlinecite{Schroder2008} are shown by the black line. 
     The green, blue and pink broken lines are calculated results with no single anisotropy ($K=0$: uniform J model), easy axis type single ion anisotropy of $K=J/4$, and $K=J/2$, respectively. }
\label{fig:Fig10} 
\end{figure}

     The origin of the spin freezing in Fe30 is pointed out to be intramolecular exchange disorder by theoretical studies using Monte Carlo simulation.\cite{Schroder2010} 
     In the simulation, the distribution of the intramolecular exchange interaction $J$ is introduced to reproduce the d$M$/d$H$ curve at low temperatures which shows  a broad dip around 5 T due to 1/3 anomaly.\cite{Schroder2008}  
     For a model with uniform $J$, the broad dip is not reproduced at all as shown in Fig. 9.\cite{Schroder2008} 
     The simulation found that the exchange disorder leads to a spin freezing state and slow spin dynamics.\cite{Schroder2010}
      In the previous simulation, any anisotropy effects and/or Dzyalosinskii-Moriya interactions were not included. 
      In order to investigate the effects of single ion anisotropy $K$ (easy-axis type) for Fe$^{3+}$ spins on the d$M$/d$H$ curve, we have calculated the d$M$/d$H$ curve using the same method in Ref. \onlinecite{Schroder2010}. 
      It turns out that, although the dip becomes broader with increasing $K$ (up to the half of $J$) as shown in Fig. 9, the d$M$/d$H$ cannot be reproduced well as in the case of $``$exchange disorder$"$ shown by the red line. 
    We also found the easy-plane type anisotropy produces similar effects as in the case of $``$easy axis$"$. 
    Thus we consider that the origin of the spin freezing is not due to single ion anisotropy effects but more likely to the distribution of the exchange interactions. 
     However, we cannot rule out the possible effects due to Dzyalosinskii-Moriya interactions. 
    This is outside the scope of the present work and remains to be confirmed by detailed calculations. 
     In any case, since it is important to know the actual structure of Fe30 at low enough temperature,
     it would be very helpful to perform low temperature X-ray diffraction measurements. 

\section{Cu3 spin tube}

\begin{figure}[b]
 \includegraphics[width=6.0cm]{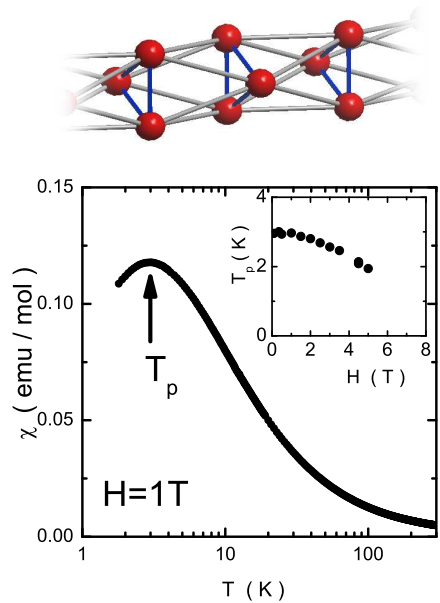} 
 \caption{(Color online) Top: Schematic structure of the triangular spin tube,  [(CuCl$_2$tachH)$_3$Cl]Cl$_2$. 
  The circles show Cu$^{2+}$ ($s=1/2$) ions. 
   The black and gray lines are corresponding to the intra-triangle exchange ($J_1\sim0.9$ K) and inter-triangle exchange ($J_2\sim1.95$ K) coupling paths.
   Bottom: Temperature dependence of the triangular spin tube measured at $H=1$ T. 
   The inset shows the $H$ dependence of the peak temperature $T_{\rm p}$. 
}
 \label{fig:Fig11}
 \end{figure}

    [(CuCl$_2$tachH)$_3$Cl]Cl$_2$ (tach=cis,trans-1,3,5-triamino-cyclohexane), is a newly synthesized triangle spin tube.\cite{Seeber2004} 
    With a trigonal crystal structure (space group P63/m, lattice parameters $a=b=12.800~\AA$, $c=12.6287~\AA$),\cite{Seeber2004} the triangles formed by three Cu$^{2+}$ (3$d^9$, $s$ = 1/2) ions are aligned to construct infinite stacks of antiprisms in a one dimensional way along the $c$ direction. 
     A schematic structure of the triangular copper chain is shown in the top part of Fig. 10. 
     Each triangle rotates 180 degree with respect to its neighbor triangle in the chain direction (a twisted triangular spin tube). 
     Each Cu spin in the triangle is coupled to two Cu spins of each neighboring triangle.　
      From the temperature dependence of the magnetic susceptibility measured between 2 and 300K, antiferromagnetic interactions of intratriangle and intertriangle are reported to be $J_1/k_{\rm B}$ $\sim$ 0.9 K and $J_2/k_{\rm B}$ $\sim$ 1.95K, respectively.\cite{Schnack2004}  
     Although initially the system was reported to have a spin singlet ground state with a spin gap $\Delta/k_{\rm B}$ $\sim$ 0.4 K from a magnetization curve at $T$ = 0.5 K,\cite{Schnack2004} theoretical investigations indicate a gapless ground state, i.e, no gapped ground state for the twisted triangular spin tube.\cite{Sakai2010, Fouet2005,Okunishi2005} 
    Experimentally, a gapless ground state of the Cu spin tube was revealed by an observation of proton NMR line broadening at very low temperatures below 1 K,\cite{Furukawa2009} which is shown below.  
   Specific heat measurements down to 100 mK also evidenced a gapless ground state of the Cu spin tube from the observed $T$ linear dependence of the specific heat below 0.6 K, which is attributed to a Tomonaga-Luttinger liquid state.\cite{Ivanov 2010}

\subsection{$^1$H NMR spectra in the spin tube}
    Figure 11(a) shows a typical temperature dependence of $^1$H NMR spectra measured at $f$ = 42.5759MHz.\cite{Furukawa2009}
    A single peak of $^1$H NMR spectrum can be observed at high temperature above $\sim$ 200 K. 
    The spectrum broadens at low temperatures and shows three peaks, denoted P1, P2 and P3, as seen in Fig. 11 (a). 
     The peak positions for P1 and P3 shift to lower and higher magnetic field, respectively, on lowering temperature. 
     On the other hand, the peak position of P2 observed around the Larmor magnetic field is independent of temperature. 
     Typical temperature dependence of NMR shift ($K$) for the P1 line is shown in Fig. 11(b). 
     Below $\sim$ 1.5K, the shifted peaks P1 and P3 disappear due to the shortening of spin-spin relaxation time $T_2$ and only the non-shifted peak P2 can be observed. 
    At very low temperatures below 0.2 K, shoulder structure can be seen on either side of the main peak. 
   T he FWHM of the P2 line as a function of temperature is shown in Fig. 11 (c), which scales well with $T$ dependence of $M/H$ above 2 K. 
     Below 2 K, the FWHM shows a local minimum around 1 K and levels off on lowering temperature. 
    Since the line width relates to the magnetization $M$, these experimental results indicate $M$ is almost constant and has finite value at very low temperatures, which is consistent with a gapless ground state of the system reported from specific heat measurements.

\begin{figure}[tb]
 \includegraphics[width=8.7cm]{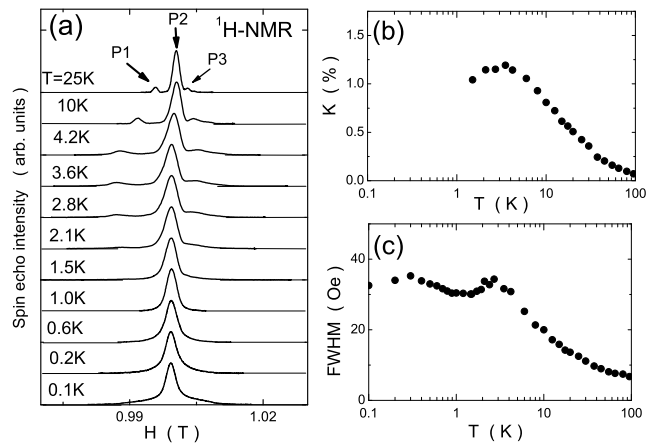} 
 \caption{(Color online) Temperature dependence of $^1$H-NMR spectra measured at $ f=42.5759$ MHz. (b) Temperature dependence of NMR shift ($K$) for the P1 line. (c)  Temperature dependence of full width of half amplitude (FWHM) for the P2 line. }
 \label{fig:Fig12}
 \end{figure}

    In general, there are two possibilities for the origin of the hyperfine field at proton sites, that is, dipolar fields and contact hyperfine fields from the Cu$^{2+}$ spins. 
    In the case of powder sample, the dipolar fields yield a broadening of the line but no net shift, while the net shift of the line can be produced by the contact hyperfine fields due to the overlap of the $s$ electron wave function at the proton sites with the $d$ electron (Cu$^{2+}$) wave function. 
    In the Cu spin tube system, protons can be mainly classified into two different sites. 
    One is the protons contributed to hydogen bonds between triangles and another is the protons in cyclohexan trimanio located between the spin tubes. 
    The almost unshifted line (P2) originates from the protons which are coupled to Cu$^{2+}$ spins with dipolar interaction only, corresponding to the protons in cyclohexan trimanio located between the spin tubes. 
    On the other hand, the shifted lines can be ascribed to the protons inside in the exchange path between Cu$^{2+}$ spins because these protons experience the isotropic hyperfine field due to the Cu$^{2+}$ spins. 
    The existence of isotropic hyperfine field for the protons is microscopic evidence of the importance of the hydrogen bonds for magnetic interactions between the triangles, responsible for the one dimensional magnetic nature of the system.

    In order to see the temperature evolution of the line width at the shoulder part of the P2 line, we define $\Delta H$ which is estimated from the line width at 10 $\%$ of the peak intensity. 
   The $\Delta H$ are displayed as a function of the temperature in Fig. 12(a) together with results measured at different magnetic fields from 0.223 T to 1.46 T. 
    The $T$ dependence of $\Delta H$  above 0.3 K are in good agreement with that of FWHM, indicative of the existence of the local minimum in the $T$ dependence of $\chi$  around 1 K. 
    It is noted that a different behavior between FWHM and $\Delta H$  can be found below 0.2K. 
    Although the origin of the different behavior is still not clear at present, it is important to point out that the increase of $\Delta H$  at low temperatures indicates the growth of a distribution of internal fields at the proton sites from the spin moments on the Cu$^{2+}$ ions. 
   Since we observed  the increase of $\Delta H$  even at a low magnetic field of $H=0.22$ T, we conclude that spin moments exist at low temperatures and the ground state of the system is not a spin singlet state but a gapless state. 
    As shown in Fig. 12 (b),  the $\Delta H$ at  $T$ = 0.1 and 1.5 K increases linearly with $H$, again consistent with a gapless ground state. 
    Our conclusion is consistent with theoretical results\cite{Sakai2010, Fouet2005,Okunishi2005} where the ground state of the twisted Cu spin tube is predicted to be not gapped but gapless.

   Fig. 12(c) shows the DC magnetization curve at $T$ = 0.09, 0.3, 0.5 and 1.8 K. 
   The magnetization has a small fraction which shows Brillouin function-like behavior in the low magnetic field region at low temperatures. 
    This is attributed to spin contributions due to magnetic impurities. 
    The fraction of the excess magnetization is estimated to be $\sim$ 2 $\%$  of the total magnetization. 
    The most important finding in the DC magnetization measurement is an obvious change in slope of the magnetization curves around 5 T below and above 0.5 K, which can be also seen in change of d$M$/d$H$ as shown in the inset of Fig. 12(c). 
   This signature can be attributed to the so-called 1/3 plateau in magnetization, since the magnetization around 5 T is $\sim$ 1 $\mu_{\rm B}$/mol, which is one-third of an expected total magnetization of 3 $\mu_{\rm B}$/mol and 5 T is close to one-third of the saturation magnetic field of 14 T.\cite{Schnack2004} 

     A phase diagram for a twisted spin tube as a function of $J_2/J_1$ and of magnetic field, $H$, has been proposed theoretically by Fouet ${\it et~al.}$,\cite{Fouet2005}, using a density matrix renormalisation group calculation. 
    When $J_2/J_1$ is smaller than 1.22, the system has a gapped ground state with a two-fold-degenerate state, due to spin chirality. 
   On the other hand, if $J_2/J_1$ is large enough, the system is considered an effective $S=3/2$ chain with a gapless ground state. 
   The phase transition between the gapped and gapless states takes place for a critical point at $J_2/J_1$ $\sim$ 1.22. 
    The 1/3 plateau in the magnetization curve is proposed below $J_2/J_1$ $\sim$  1.6. 
   Although the ratio, $J_2/J_1$, is reported previously to be $\sim$ 2.16 from the temperature dependence of magnetic susceptibility, the observation of  1/3 plateau with the gapless ground state indicates that the ratio should fall in the interval 1.22 $<$ $J_2/J_1$ $<$ 1.6, suggesting that the Cu3 spin tube is located close to the critical point of the phase transition.

\begin{figure}[tb]
 \includegraphics[width=9.0cm]{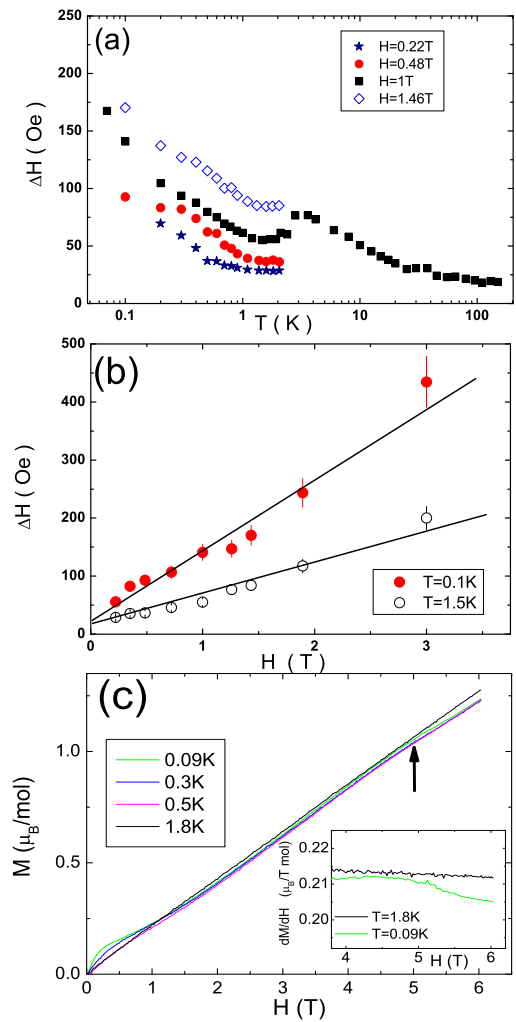} 
 \caption{(Color online) $T$ dependence of $\Delta$$H$ for the P2 line under various magnetic fields. 
    (b) External field dependence of $\Delta H$ at $T$ = 0.1 K and 1.5 K.  
 (c) $H$ dependence of DC magnetization $M$  at $T$ = 0.09, 0.3, 0.5 and 1.8 K. 
    The arrow shows  the magnetic field,  where the slope in magnetization changes below 0.5 K. 
     The inset shows d$M$/d$H$ for $T$ = 1.8 K and 0.09 K around 5 T. 
}
 \label{fig:Fig13}
 \end{figure}

\subsection{Spin dynamics in the spin tube}

    Now let us discuss dynamical properties of the Cu$^{2+}$ spins. 
     We have carried out proton $T_1$ measurements at the peak position of the P2 line in a wide temperature range of $T=0.05-300$ K.\cite{Furukawa2011}
     Fig. 13 shows the temperature dependence of 1$/T_1$ under various magnetic fields. 
    With decreasing temperature from room temperature, 1$/T_1$ decreases smoothly and starts to increase around 2K, then shows a peak around 0.6 K. 
   As the external field is increased, the peak temperature of 1$/T_1$ shifts to higher temperature and at the same time the height of 1$/T_1$ becomes smaller. 
    At very low temperatures below $\sim$ 0.1 K, 1$/T_1$ seems to merge at different magnetic fields, which is probably due to experimental error due to heat up problems for the $T_1$ measurements. 
   The inset of Fig. 13 shows the $H$ dependence of 1$/T_1$ measured at $T=1.5$ K where 1$/T_1$ is found to be proportional to 1/$H^{0.5}$ whose behavior is well explained by spin diffusion effects in one dimensional magnetic systems.\cite{Borsa1974} 
    This confirms again the one dimensionality of the twisted triangular spin tube. 

 \begin{figure}[tb]
 \includegraphics[width=6.0cm]{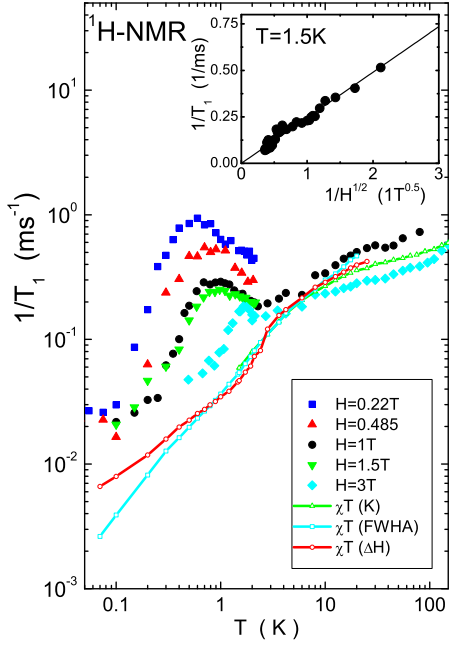} 
 \caption{(Color online) Temperature dependence of 1/$T_1$ under various magnetic fields. 
    Temperature dependences of $\chi T$ estimated from $K$, FWHM and $\Delta$ shown by open symbols are plotted with arbitrary units in vertical axis to show the same temperature dependence of 1/$T_1$ at high temperatures. 
      The inset shows the $H$ dependence of 1/$T_1$ as a function of 1/$H^{0.5}$ at $T=1.5$ K. 
     The solid line in the inset shows a relation of 1/$T_1$ $\propto$ 1/$H^{0.5}$  expected from spin diffusion effects in one dimensional spin systems. }
 \label{fig:Fig14}
 \end{figure}

     We analyze the 1/$T_1$ data with the same equation used for the case of Fe30 spin ball.
    As shown in Eq. (6), if the $\Gamma$ (here  $\Gamma$ corresponds to the inverse of the correlation time of the fluctuating hyperfine fields at the H sites, due to the Cu$^{2+}$ spins) is independent of temperature, temperature dependence of 1$/T_1$ is simply expressed by $\chi T$. 

     The open symbols (circles, squares and triangles) in Fig. 13 show the $T$ dependences of $\chi T$ estimated from the $T$ dependences of $K$, FWHM and $\Delta H$ of $^1$H NMR experiment. 
    As can be seen in the figure, the $T$ dependence of 1$/T_1$ seems to be in good agreement with that of $\chi T$, although 1$/T_1$ depends on the external field due to the spin diffusion effects as described above. 
   Thus the nuclear spin relaxations in the twisted spin tube above $\sim$ 3 K is mainly explained by the paramagnetic fluctuations of the Cu$^{2+}$ spins. 
  On the other hand, the simple paramagnetic fluctuations model can not reproduce the experimental results below 3K, especially the peak behavior in 1$/T_1$.
    It is noted that the $H$ dependence of 1$/T_1$ above the peak is different from that of 1$/T_1$ below the peaks. 
    The $H$ dependence above the peak temperature can be explained by the spin diffusion nature as discussed above, but 1$/T_1$ depends on $H$ stronger than the $H$ dependence of spin diffusion effects below the peak temperature. 

      In order to analyze the $T$ and $H$ dependences of 1$/T_1$ at low temperatures below 3 K using Eq.  (6), we re-plotted 
the 1/$T_1T\chi$ vs. $T$ as in the case of Fe30, as shown in Fig. 14(a) where the $\chi T$ values used are estimated from $T$ dependence of $\Delta H$.   
    Although there is a slight difference between $\chi T$ estimated from FWHM and $\Delta H$ at very low temperatures, it is only at very low temperatures below $\sim$ 0.2 K, so that our analysis is not modified very much.  
     Using $A=1.96\times10^{12}$ (rad$^2$$\cdot$K$\cdot$emu/mol/s$^2$) and $\Gamma=3.48\times10^8T^2$  (rad/s), the experimental results are well reproduced by Eq. (6) as shown in Fig. 14(a) by solid lines for different magnetic fields. 
     These results indicate that the peak behavior observed in the $T$ dependence of 1/$T_1T\chi$ originates from a crossover between the fast-motion regime and the slow-motion regime, and fluctuation frequency of Cu$^{2+}$ spins below the peak is less than the NMR frequency at low temperatures. 

\begin{figure}[tb]
 \includegraphics[width=8.0cm]{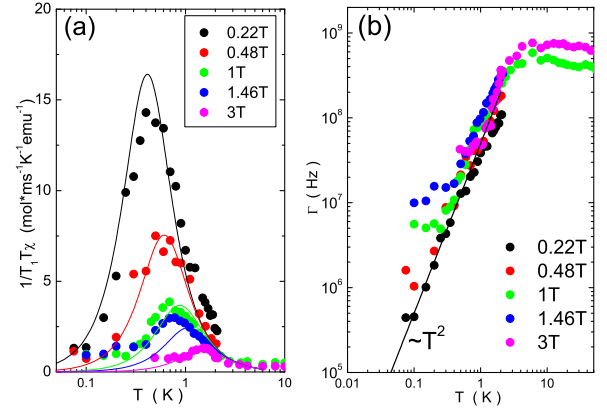} 
 \caption{(Color online) ) Temperature dependence of 1/$T_1T\chi$  under various magnetic fields. 
    The solid lines are calculated results using Eq. (2). 
     (b) Temperature dependence of fluctuation frequency ($\Gamma$) of the Cu$^{2+}$ spins estimated from the $T$ dependence of 1/$T_1T\chi$. 
   The solid line in the figure corresponds to a relation of $\Gamma \propto T^2$.    }
 \label{fig:Fig15}
 \end{figure}

    We extract the temperature dependence of $\Gamma$ from the temperature dependence of 1/$T_1T\chi$  assuming Eq. (6) is valid for the entire temperature region. 
    The estimated $T$ dependences of $\Gamma$ for different external magnetic fields are shown in Fig. 14(b). 
    This log-log plot shows very clearly that $\Gamma$ has power low behavior ($T^{\sim 2}$) at low temperatures below $\sim$ 3 K and shows an almost constant value of the order of 10$^9$ Hz at high temperatures. 
     Thus we conclude that the fluctuation frequency of the Cu$^{2+}$ spins is almost constant at high temperatures and becomes slower and slower with a relation of $T^2$ on lowering the temperature below 3 K.
      At low enough temperature, Cu$^{2+}$ spins can fluctuate very slowly,  less than NMR frequency ( $\sim$ 10$^6$ Hz). 
     Such a spin freezing phenomena could explain the different behavior in $\Delta H$ and FWHM at low temperatures. 
    The protons close to one Cu$^{2+}$ spin experience non-vanishing dipolar fields from the spins, which leads to broadening of the spectrum. 
     On the other hand, the dipolar fields at proton sites located at the middle position between the antiferromagnetically coupled Cu spins could be canceled out, resulting in sharper line width. 
     Thus the slight decrease of the FWHM below 0.2 K is though to be a signature of growth of antiferromagnetic correlations between the Cu$^{2+}$ spins in the slow-motion regime.  
  
  　 In usual $s=1/2$ one dimensional quantum spin systems with a gapless ground state, 1$/T_1$ is considered to be in the fast-motion regime because such a spin freezing is not expected due to a continuous energy dispersion for the spin excitation spectrum. 1$/T_1$ shows temperature independent behavior at high temperatures and a logarithmic increase at low temperatures without any crossover between fast-motion and slow-motion regimes.\cite{Takigawa1996}  
     This is in sharp contrast to the present case. 
     On the other hand, a similar crossover between slow-motion and fast motion regimes was observed in various nanoscale magnetic molecules.\cite{Borsa2006, Schroder2010, Furukawa2012, Furukawa2000, Furukawa2001, Kubo2002, Goto2003, Morello2003, Furukawa20012, Ueda2002, Furukawa2003, Baek2005,Baek2004} 
    Based on a similar analysis, the temperature dependence of $\Gamma$ in the magnetic molecules is found to be proportional to $T^{3.5}$. This $T$ dependence is explained by the spin-phonon interactions.\cite{Rousochatzakis2009,Santini2005} 
    On the other hand, $\Gamma$ $\sim$ $T^2$ behavior in the present system cannot be explained by the spin-phonon interactions. 

    Recently the specific heat of the system has been measured down to 0.1 K,\cite{Ivanov 2010} confirming no three dimensional magnetic ordering under zero magnetic field down to 0.1 K. 
     The temperature dependence of specific heat shows a broad peak around 2 K and rapid decrease below the peak with decreasing  temperature.
     Below 0.5 K, the specific heat is almost proportional to $T$, which indicates a gapless ground state of the system. 
    This is consistent with our experimental results. Interestingly, $T$ linear behavior observed in the specific heat is one of the signatures of the Tomonaga-Luttinger liquid (TLL) state. 
    In the case of TLL state, 1$/T_1$ is expected to follow the relation of 1$/T_1$ $\sim$ 1/$T^{\alpha}$ with $\alpha = 0.5 - 0.66$.\cite{Goto2006} 
    Goto et al. reported a observation of 1/$T_1$ $\sim$ 1/$T^{0.6}$ behavior in one dimensional $s=1$ quantum spin system (CH$_3$)$_4$NNi(NO$_2$)$_3$ in a gapless state under magnetic fields, suggesting the realization of TLL state in that system.\cite{Goto2006} 
     In the present system, 1$/T_1$ does not show such a behavior expected for the TLL states. 
     Thus it seems to be difficult to explain the anomalous behavior of 1$/T_1$ observed in the twisted Heisenberg triangular spin tube by the conventional explanations described above. 
    At present, although it is not clear the reasons for the slowing down of spin fluctuations at low temperatures, it would be interesting if these peculiar behaviors observed in the twisted Cu spin tube were related to spin chirality.

\section {Summary}
      A review of NMR studies performed on the three nanoscale molecular magnets with peculiarly-structured frustrated systems has been presented. 

   (i) V15 : a model system of  $s$=1/2 Heisenberg triangular antiferromagnet. 
    The $^{51}$V NMR spectrum measurements  at very low temperatures below 0.1 K  and as a function of the external applied field have allowed the determination of the local microscopic spin configuration in the frustrated ground state. 
    It is found that the pure chiral states described by  Eq. (3) are broken down in the V15 nanomagnet. 
   The two quasidegenerate $S_{\rm T}$ = 1/2 ground states correspond to two different spin configurations 
which can be described by the two eigenfunctions $\psi_{\rm \alpha}$ and $\psi_{\rm \beta}$. 
   We also found that the lower energy of the two $S_{\rm T}$ = 1/2 split states is the one pertaining 
to the eigenfunction  $\psi_{\rm \alpha}$. 
   The local spin configuration is consistent with a small structural distortion from an equilateral 
triangle to a nearly isosceles one. 
       The dynamical magnetic properties of V15 were investigated by  proton spin-lattice relaxation rate (1/$T_1$) measurements. 
    In the $S_{\rm T}$ = 3/2 state,   1/$T_1$ shows thermally activated behavior as a function of temperature. 
    On the other hand, a temperature independent behavior of 1/$T_1$ at very low temperatures is observed in frustrated $S_{\rm T}$ = 1/2  ground state below 2.7 Tesla. 
    Possible origins for the peculiar behavior of 1/$T_1$ have been discussed in terms of magnetic fluctuations due to spin frustrations.

(ii) Fe30 : a spin ball. 
    Static and dynamical properties of Fe$^{3+}$ ($s = 5/2$) have been investigated by proton NMR spectra and 1/$T_1$ measurements. 
   From the temperature dependence of 1/$T_1$, the fluctuation frequency of the Fe$^{3+}$ spins is found to decrease with decreasing temperature, indicating spin freezing at low temperatures. 
  The spin freezing is also evidenced by the observation of a sudden broadening of $^1$H NMR spectra below 0.6 K. 
  The origin of the spin freezing in Fe30 has been suggested to be the distribution of the exchange interactions. 

(iii) Cu3 : a twisted spin tube.
          $^1$H NMR and DC magnetization data in the Cu spin tube were described. 
  An observation of magnetic broadening of $^1$H NMR spectra at low temperatures below 1 K directly revealed a gapless ground state of the system. 
   The DC magnetization data below 1 K are also consistent with the gapless ground state. 
   We also found the clear change in the slope of the magnetization curves around 5 T attributed to the so-called 1/3 plateau in magnetization.    
    The temperature dependence of $1/T_1$ under various magnetic fields revealed an usual slow spin dynamics in the Cu3 spin tube.
    The temperature dependence of 1/$T_1$ was not simply explained by the spin-phonon interactions model, nor is it the Tomonaga-Luttinger liquid model. 
     Further theoretical studies are required to understand the peculiar behavior of 1/$T_1$ in the quasi one dimensional Cu3 spin tube.      
     It is important to point out that NMR appears to be a unique tool for determination 
of the local spin configuration, ground state properties and low energy spin dynamics of frustrated systems.

\begin{acknowledgments}
    The author would like to acknowledge collaborations and fruitful discussions with Ferdinando Borsa, Ken-ichi Kumagai,  Alessandro Lasscialfari, Yusuke Nishisaka, Yuri Fujiyoshi, Yuzuru Sumida, Yuichi Hatanaka,  Edoardo Micotti, Marshall Luban, Ruslan Prozorov, Christian Schr\"oder, J\"urgen Schnack, and Hiroyuki Nojiri. 
    Special thanks go to Paul K\"ogerler, Xikui Fang, and Leroy Cronin who have been succeeded in synthesizing valuable crystals which makes it possible for us to carry out this research. 
    Monte Carlo simulation for the Fe30 magnetization has been performed by Christian Schr\"oder.
    DC magnetization for the Cu spin tube at low temperatures below 1 K was measured by Hiroshi Amitsuka, Ken-ichi Tenya, and Yusei Shimizu at Hokkaido University, Japan.
     A part of NMR work has been performed at Hokkaido University in Japan, which is supported by 21$^{\rm st}$ Century COE Programs "Topological Science and 
  Technology" at Hokkaido University and Grant-in-Aid from the Ministry of Education, Culture, Sports, Science and 
 Technology of Japan. 
     The work at Ames Laboratory was supported by the U.S. Department of Energy (DOE), Office of Basic Energy Sciences, Division of Materials Sciences and Engineering. Ames Laboratory is operated for the U.S. DOE by Iowa State University under Contract No. DE-AC02-07CH11358.
\end{acknowledgments}

\end{document}